\documentclass[letter]{acoust}

\def\textfraction{.008}

\def\dbltopfraction{.99}
\def\dblfloatpagefraction{.99}

\makeatletter
\long\def\@makefntext#1{\vskip3\p@ \hsize\columnwidth \par \noindent \footnotesize \hskip6\p@ $^{\@thefnmark}$\hskip1\p@#1\vskip-3\p@}
\makeatother

\def\header{Safeguarding sounds for measurement}

\title{Safeguarding test signals for acoustic measurement using arbitrary sounds}
\subtitle{Measuring impulse response by playing music}

\author{Hideki Kawahara$^{1}$\thanks{kawahara@wakayama-u.ac.jp}, Kohei Yatabe$^{2}$\thanks{k.yatabe@asagi.waseda.jp}}

\inst{$^{1}$Wakayama University, 930 Sakaedani, Wakayama, 640-8510 Japan \\
$^{2}$Waseda University, 3-4-1 Ookubo, Shinjuku-ku, Tokyo, 169-8555 Japan}

\kword{discrete Fourier transformation, periodic signal, impulse response, spurious responses, entertaining test sounds, loop music}


\mladdress{〒630-0121 生駒市北大和２−１４−４\\ 河原英紀} 

\class{5} 

\recommend{2}  

\begin{document}
\maketitle
\renewcommand{\thefootnote}{\arabic{footnote}}

\section{Introduction}

We propose a simple method to measure acoustic responses using any sounds by converting them suitable for measurement.
This method enables us to use music pieces for measuring acoustic conditions.
It is advantageous to measure such conditions without annoying test sounds to listeners.
In addition, applying the underlying idea of simultaneous measurement of multiple-paths\cite{kawahara2020apsipa,kawahara2021icassp} provides practically valuable features. 
For example, it is possible to measure deviations (temporally stable, random, and time-varying) and the impulse response while reproducing slightly modified contents under target conditions.
The key idea of the proposed method is to add relatively small deterministic signals that sound like noise to the original sounds.
We call the converted sounds as \textit{safeguarded test signals}.


\section{Safeguarded transfer function measurement}
Let $x[n]$ be a periodic discrete-time signal with a period $L$.
Convolution of $x[n]$ and the impulse response $h[n]$ of the target system yields the output $y[n]$.
Because the signal is periodic, the DFT (Discrete Fourier transform) of $x[n]$ and $y[n]$ segments (their length is $L$) are invariant other than the phase rotation proportional to frequency.
Let $X[k]$ and $Y[k]$ represent their DFT, where $k$, ($k = 0, \ldots, L-1$), is the discrete frequency.
Then, the ratio $Y[k]/X[k]$ is independent of the location of the segment.
This ratio agrees with the DFT $H[k]$ of the impulse response $h[n]$,
where $X[k] \neq 0$ for all $k$ values is the condition of this relation to provide physically meaningful results.

However, this simple solution is sensitive to noise when the absolute value $\left|X[k]\right|$ is very small relative to absolute values $\left|H[k]\right|$ of other $k$ values.
We propose to limit the absolute value $\left|X[k]\right|$ to have larger value than a threshold\footnote{Generally the threshold is a function of discrete frequency $\theta_L[k]$. We use a constant value here to make explanations simple.}.
We use the following equation to derive the DFT $X_\mathrm{s}[m]$ of the safeguarded signal $\tilde{x}_\mathrm{s}[n]$.
\begin{align}
X_\mathrm{s}[k] & = \left\{\begin{array}{lll}
\displaystyle \frac{\theta_LX[k]}{\left|X[k]\right|} \ \ & \mbox{for} \ \ & 0 < \left|X[k]\right| < \theta_L  \\
X[k] & { } & \theta_L \leq \left|X[k]\right| 
\end{array}\right. ,
\end{align}
where we set $X_\mathrm{s}[k] = \theta_L$ when $X[k] = 0$.
Then, we derive the safeguarded transfer function $H_\mathrm{s}[k]$ as follows.
\begin{align}
H_\mathrm{s}[k] & = \frac{Y_\mathrm{s}[k]}{X_\mathrm{s}[k]} ,
\end{align}
where $Y_\mathrm{s}[k]$ represents the DFT of the output of the target system for periodic test signal $\tilde{x}_\mathrm{s}[n]$. 
Because the safeguarded signal $\tilde{x}_\mathrm{s}[n]$ is periodic, we can make \textit{the safeguarded test signal} for acoustic measurement by concatinating it as many times as required.
For analyzing the safeguarded transfer function, we can select the safeguarded test and the output segment anywhere, obeying one rule.
The segment has to have the length of exactly $L$ samples\footnote{Do not use the initial segment of length $L$ samples (plus samples for propagation delay from the sound source to the microphone), because it does not have the preceding cycle.}.

\section{Measurement of other responses}
When the target system is an LTI-system (LTI: liner time-invariant), and no observation noise exists, the calculated safeguarded transfer function $H_\mathrm{s}[k]$ is identical, irrespective of the location of the safeguarded test and the output segments.
Also, $H_\mathrm{s}[k]$ is independent of the used safeguarded test signals.
However, it is not the case in measuring acoustic systems in the real world.
We can use these differences of $H_\mathrm{s}[k]$ measured at different observation locations and using different safeguarded test signals to separate the LTI-response and other spurious responses. 
They are signal-induced deterministic responses and random responses\footnote{In addition to background noise and observation noise (they are source independent), there are source-related random noises. 
For example, they are; turbulent noise caused by strong low-frequency airflow in the bass-reflex port and high-frequency phase modulation noise due to the doppler effect caused by air move\cite{kawahara2020apsipa}.}.

\subsection{Separation of random responses}
An additive noise $d[n]$ in output observation produces a deviation term $D[k]$.
We define the time-invariant response $H_\mathrm{sTI}[k]$ and the squared absolute random response $|D_\mathrm{sTV}[k]|^2$ by measuring the system $M$ times.
\begin{align}
H_\mathrm{sTI}[k] & = \frac{1}{M}\sum_{m=1}^{M} H_\mathrm{s}^{\{n_m\}}[k] \\
|D_\mathrm{sTV}[k]|^2 & = \frac{1}{M-1}\sum_{m=1}^{M} \left| H_\mathrm{s}^{\{n_m\}}[k]- H_\mathrm{sTI}[k] \right|^2 ,\label{eq:randESt}
\end{align}
where use of superscript $a^{\{n_m\}}$ represents that the calculation of $a$ uses the segment starting from the discrete-time $n_m$\footnote{It is better not to overlap analysis segments.}.

\subsection{Separation of signal-dependent responses}
For LTI-systems, the time-invariant response $H_\mathrm{sTI}[k]$ is identical, irrespective of the test signals.
However, again, it is not the case.
We define the linear time-invariant response $H_\mathrm{sLTI}[k]$ by averaging responses measured using different test signals. 
Then, we represent the squared absolute signal-dependent responses $|H_\mathrm{sSDR}[k]|^2$ by averaging the squared absolute valued deviation from the time-invariant response.
We use an index variable $p$ to identify the member of the set of test signals $\Omega_\mathrm{P} = \{ \tilde{x}_\mathrm{s}^{\{p\}}[n] \ |\  p = 1, \ldots, P \}$.
\begin{align}
H_\mathrm{sLTI}[k] & = \frac{1}{P} \sum_{p=1}^{P} H_\mathrm{s}^{(p)}[k] \\
|H_\mathrm{sSDR}[k]|^2 & = \frac{1}{P-1} \sum_{p=1}^{P}\left|H_\mathrm{s}^{(p)}[k] - H_\mathrm{sLTI}[k] \right|^2 , \label{eq:sigDepDev}
\end{align}
where $a^{(p)}$ represents that we used the $p$-th member of the set $\Omega_\mathrm{P}$ to calculate $a$.


\section{Numerical simulation}
We conducted a set of numerical simulations to check the feasibility of the proposed method.
We used white noise for the original test signal and studied the effect of safeguarding by flooring the low-level absolute values.

\subsection{Effect of flooring in LTI-response}
\begin{figure}[tb]
\begin{center}
\includegraphics[width=\hsize]{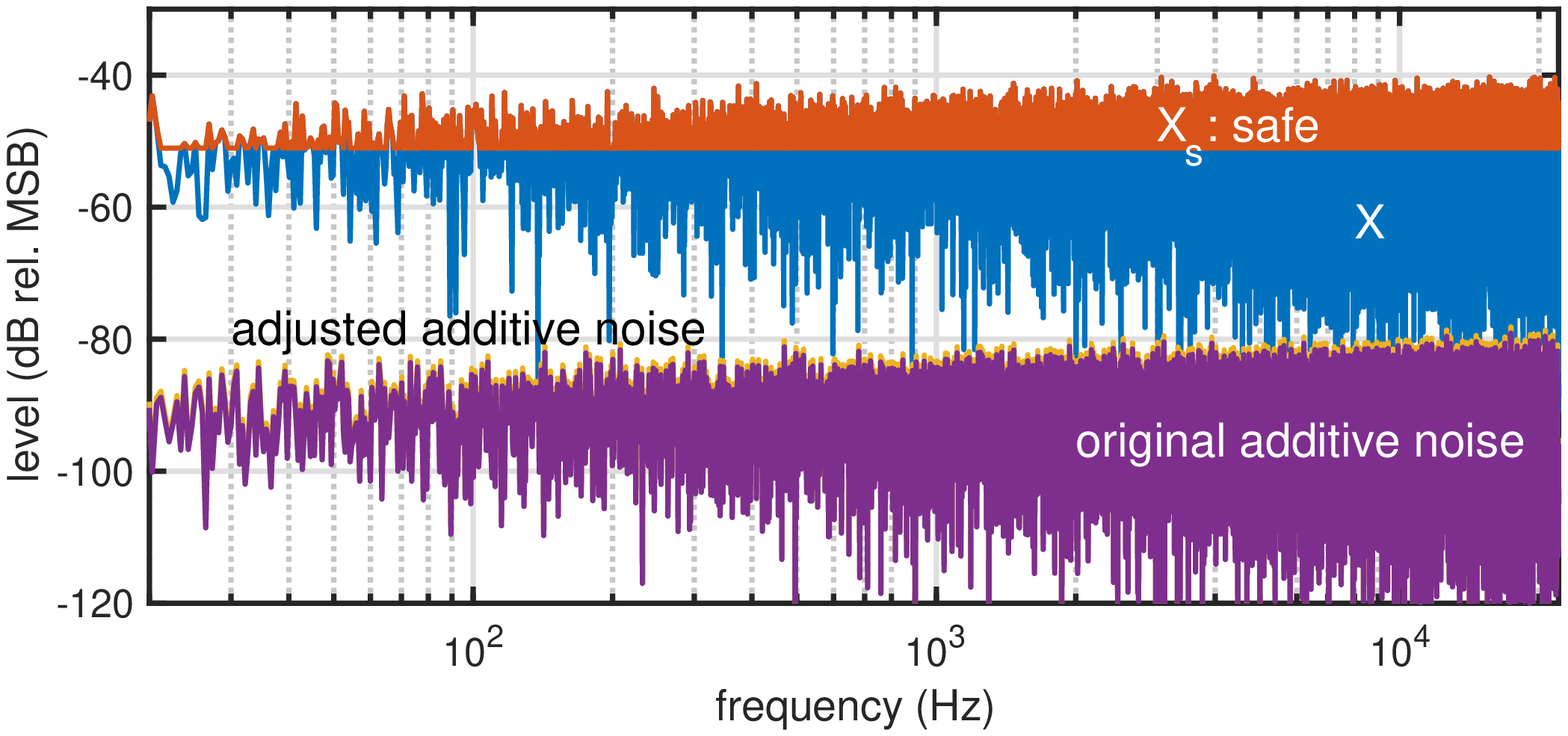}\\
\includegraphics[width=\hsize]{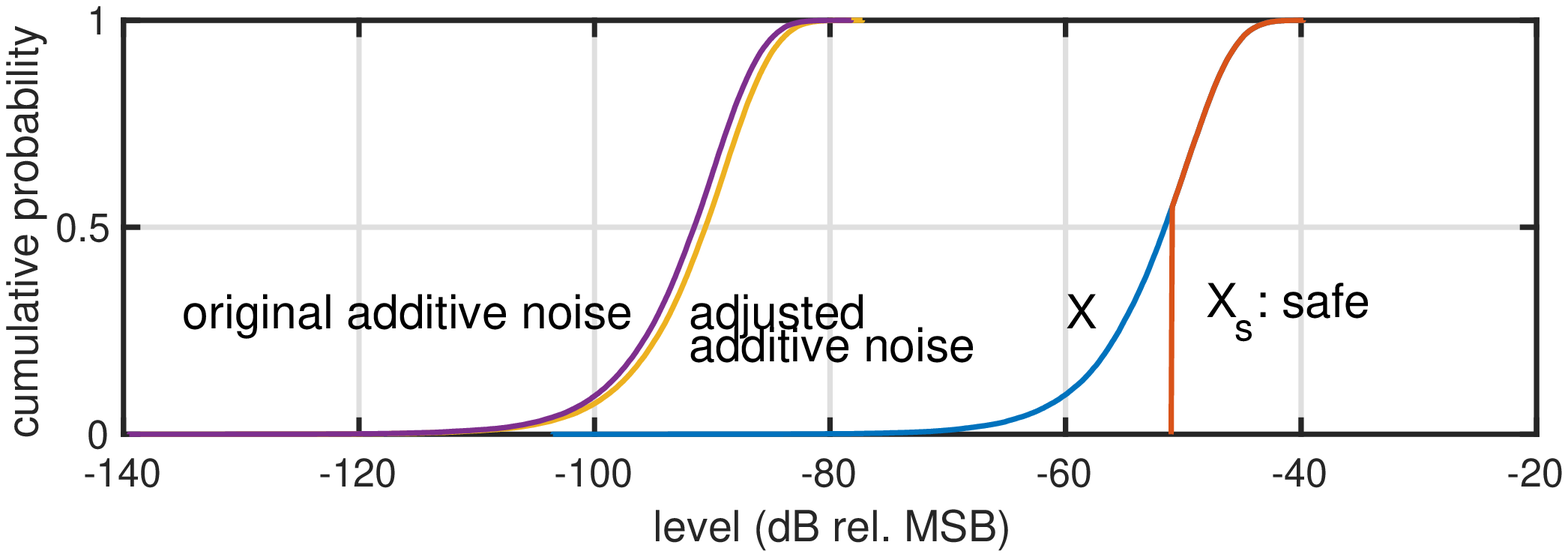}\\
\includegraphics[width=\hsize]{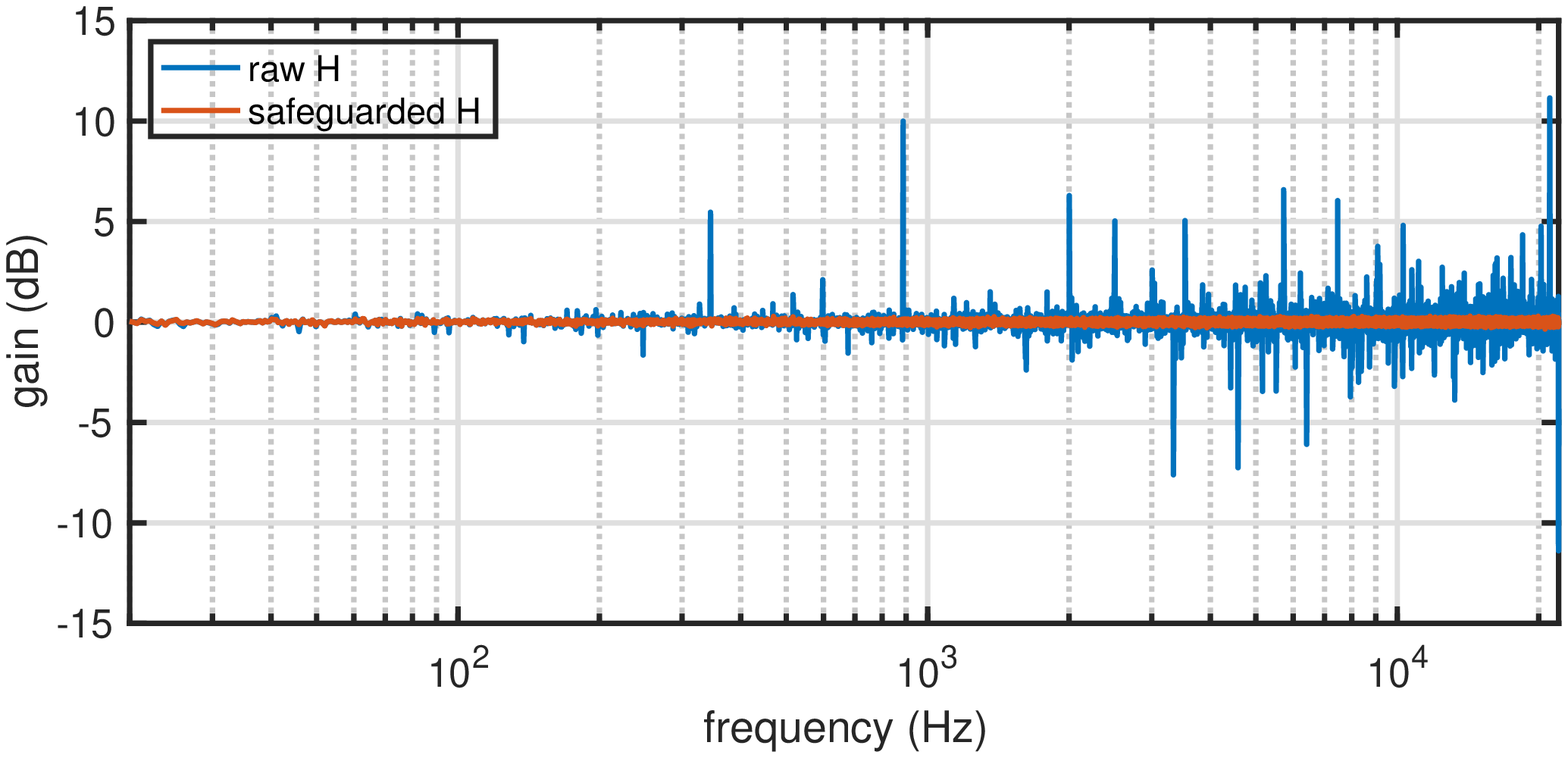}\\
\end{center}
\vspace{-3mm}
\caption{The top panel shows power spectra of the original white noise, safeguarded noise, and the background white noise (original and adjusted).
The middle panel shows the level distribution. 
The bottom panel shows the estimated gains using the original signal and the safeguarded signal.}
\vspace{-2mm}
\label{fig:whiteNoiseClipSpec}
\end{figure}
Figure~\ref{fig:whiteNoiseClipSpec} shows simulation results using additive white noise with 40~dB SNR.
We adjusted the noise level for the safeguarded signal because flooring increases the signal power.
The threshold $\theta_L$ value we used here is the average absolute value of the original spectrum.
The sampling frequency was 44100~Hz, and the signal length was 100000 samples.

\begin{figure}[tb]
\begin{center}
\includegraphics[width=\hsize]{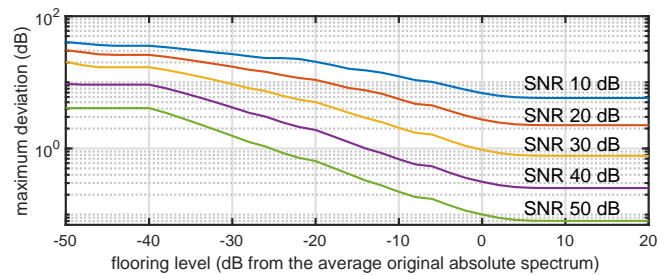}\\
\end{center}
\vspace{-2mm}
\caption{Effect of flooring on the maximum deviations from the ground truth (0~dB) for different SNR conditions. Flooring level -50~dB does not change any frequency bins and 20~dB changes all bins.}
\vspace{-2mm}
\label{fig:clipOnMaxDev}
\end{figure}
Figure~\ref{fig:clipOnMaxDev} shows the maximum deviation of the safeguarded gain function for different SNR settings.
Flooring significantly reduces maximum deviations.

Safeguarding by flooring adds a deterministic signal that sounds like noise.
Regression analysis of the flooring level $\theta_L|_\mathrm{dB}$ (represented in dB) and the power of the deterministic signal ($\sigma_\mathrm{dB}$: also represented in dB) resulted in the following experimental relation.
\begin{align}
    \sigma_\mathrm{dB} \sim -10.321 + 1.995 \ \theta_L|_\mathrm{dB} ,
\end{align}
where the intercept and the slope indicates that 0~dB flooring adds 10~dB smaller noise and the level decreases two times faster than the flooring level decrease.
For example, the -10~dB flooring level makes SNR 30~dB.
This relation suggests that safeguarding does not severely damage the quality of the original signal (for example, a music piece).

\begin{figure}[tb]
\begin{center}
\includegraphics[width=\hsize]{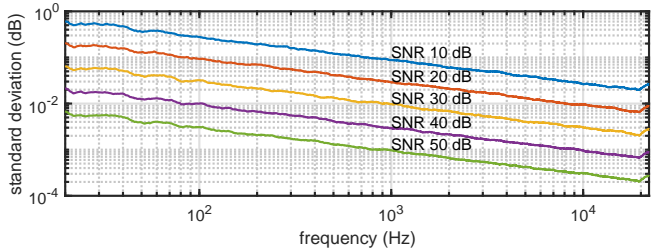}\\
\end{center}
\vspace{-2mm}
\caption{Standard deviations of the smoothed gain estimated using safeguarded signals. The flooring level is 0~dB and the smoothing width is one-third octave.}
\vspace{-2mm}
\label{fig:whiteNSmoothingSD}
\end{figure}
The estimated gain using safeguarded signal still has random peaks and dips.
Spectral smoothing is a common practice to make spectral characteristics of acoustical systems.
Figure~\ref{fig:whiteNSmoothingSD} shows the standard deviation of the smoothed gain functions using one-third octave width rectangular smoothing.
Comparison with Fig.~\ref{fig:clipOnMaxDev} illustrates that spectral smoothing significantly reduces deviations.

\subsection{Random response}
Measuring many times using the same test signal provides the random response estimate by using Eq.~\ref{eq:randESt}.
\begin{figure}[tb]
\begin{center}
\includegraphics[width=\hsize]{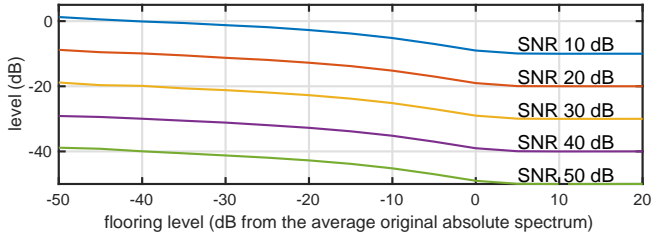}\\
\end{center}
\vspace{-2mm}
\caption{Estimated rondom response level dependency on the flooring level.}
\vspace{-2mm}
\label{fig:randomLevelMod}
\end{figure}
Figure~\ref{fig:randomLevelMod} shows the estimated random response level.
Note that the estimate at 20~dB flooring provides the correct estimate of the noise level because the test signal has a constant absolute value with randomized phase; in other words, it is a periodic pseudo-random noise.\footnote{CAPRICEP and FVN also provide correct estimate\cite{kawahara2020apsipa,kawahara2021icassp}.}

\subsection{Signal-dependent response}
Measuring many times using the same test signal does not provide signal-dependent deviations such as harmonic distortion and intermodulation distortion caused by nonlinearity. 
We introduced an exsample non-linearity using the following equation.
\begin{align}
y = \frac{1}{\alpha} \left(\exp(\alpha x) - 1 \right) .
\end{align}

We use Gaussian white noise as $x$ and added the other Gaussian noise to the output $y$ with the given SNR.
We prepared four types of the original segments and repeated each segment four times, assuming a piece of typical loop music having prospective application in mind.

\begin{figure}[tb]
\begin{center}
\includegraphics[width=\hsize]{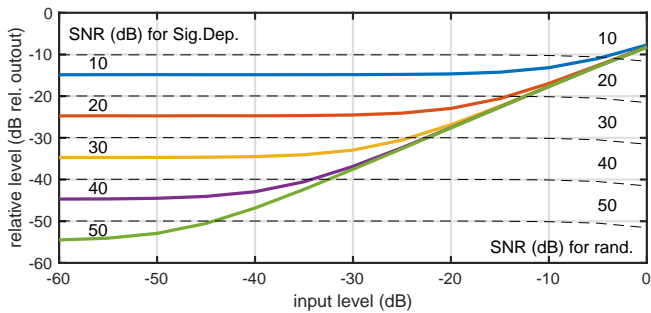}
\end{center}
\caption{Input level dependency of the random response and the signal dependent response ($\alpha = 0.4$).}
\label{fig:nlOnDeviation}
\end{figure}
Figure~\ref{fig:nlOnDeviation} shows the results.
Colored lines represent the signal-dependent response level calculated using Eq.~\ref{eq:sigDepDev}.
The black dashed lines represent the random response level calculated using Eq.~\ref{eq:randESt}.
The random response levels normalized by the total output levels are virtually constant reflecting the assigned SNR.
The signal-dependent levels decreases as the input level decreases.
This decriment satulates at the effective (-3~dB for each doubling of number of repetitions) random response level.

\section{Example measurement using a loop music}
We conducted example acoustic measurements in a connected Japanese room with an area of about 40~m$^2$.
The sound source is a loudspeaker (Fostes FF85WK) with a bass-reflex enclosure (Fostex BK85WB 2).
To drive the loudspeaker, we used a power amplifier (Fostex AP20d) connected to an audio interface (PRESONUS STUDIO $2|6$).
A wide-range omnidirectional condenser microphone (EARTHWORKS M50) connected to the audio interface acquired reproduced sounds.

We used four loop-music (5~s each) pieces composed for this research.
We mixed the stereo track into a monaural track.
The sampling frequency was 44100Hz.

\subsection{Example-1: single test segment}
For the first experiment, we located the microphone at 10~cm in front of the center of the loudspeaker.
The sound pressure level, measured using A-weighting, at the microphone was 95.8~dB.
The background noise level was 24~dB.
We repeated each segment six times.
We used the middle four segments for calculation\footnote{For loop music, it is better to repeat each segment four times and use three segments excerpted after expected reverberation time plus propagation delay. As far as the length of the segment is exactly identical to the length of the signal period, it is not necessary to match the location with the beginning of the segment.}.

\begin{figure}[tb]
\begin{center}
\includegraphics[width=\hsize]{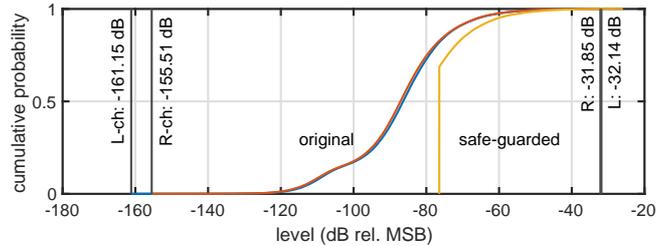}\\
\end{center}
\vspace{-2mm}
\caption{Absolute value distribution of a loop music segment. We set the threshold $\theta_\mathrm{s}$ to the average value.}
\label{fig:specLevelDistribution}
\end{figure}
Figure~\ref{fig:specLevelDistribution} shows the absolute value distribution of the DFT $X[k]$ of the original signal.
We floored absolute values at the average absolute value of the DFT. 

\begin{figure}[tb]
\begin{center}
\includegraphics[width=\hsize]{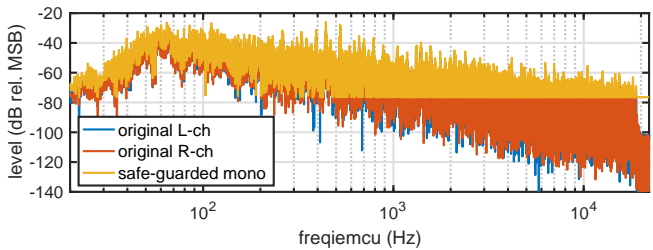}\\
\end{center}
\vspace{-2mm}
\caption{The absolute value of DFT of the original signal and the safeguarded signal.}
\label{fig:safeGuardedSpectrum}
\vspace{-2mm}
\end{figure}
Figure~\ref{fig:safeGuardedSpectrum} shows the absolute DFT values of the original signal and the safeguarded signal.
The safeguarded signal sounds like the original signal with a slight white noise.

\begin{figure}[tb]
\begin{minipage}{\hsize}
\begin{center}
\includegraphics[width=\hsize]{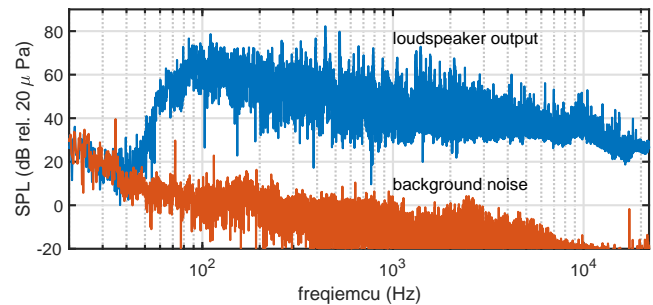}
\end{center}
\end{minipage}
\vspace{-2mm}
\caption{The DFT of the measured segment and background noise segment.}
\label{fig:measuredSigAndBg}
\vspace{-2mm}
\end{figure}
Figure~\ref{fig:measuredSigAndBg} shows calibrated sound pressure level (SPL) of DFT of the acquired loudspeaker output and the background noise.
Note that the loudspeaker output at a high-frequency range reflects spectral flooring.


\begin{figure}[tb]
\begin{minipage}{\hsize}
\begin{center}
\includegraphics[width=\hsize]{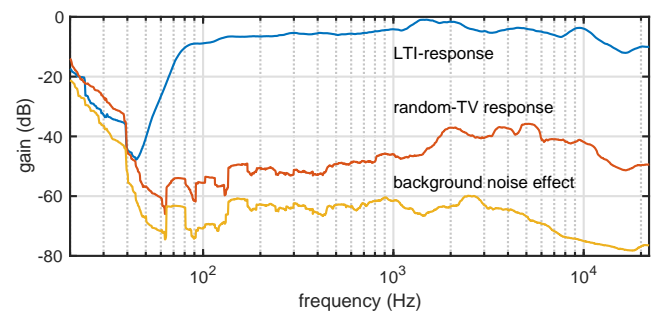}
\end{center}
\end{minipage}
\vspace{-2mm}
\caption{Separated responses with 1/3 octave smoothing.}
\label{fig:smoothedOneTypResponse}
\vspace{-2mm}
\end{figure}
Figure~\ref{fig:smoothedOneTypResponse} shows the LTI-response, random and time-varying response, and the effects of the background noise.
We used 1/3 octave smoothing to clarify essential features of the responses.
This representation suggests that the random component in the high-frequency region is the result of a high-sound pressure level.

\subsection{Example-2: multiple test segments}
\begin{figure}[tb]
\begin{center}
\includegraphics[width=\hsize]{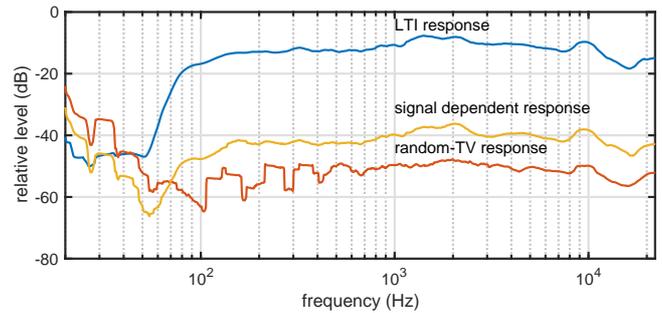}\\
\vspace{2mm}
\includegraphics[width=\hsize]{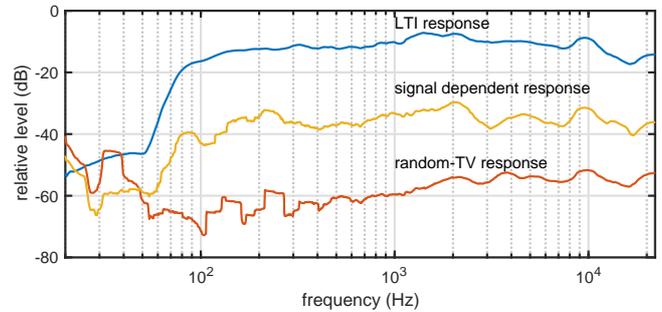}\\
\end{center}
\vspace{-2mm}
\caption{Measured responses with A-weighting SPLs 87.1~dB (upper plot) and 93.5~dB (lower plot).}
\vspace{-2mm}
\label{fig:fostex10cm}
\end{figure}
Figure~\ref{fig:fostex10cm} shows the effects of playback levels.
The upper plot shows the results of the sound pressure level at A-weighting 87.3~dB.
The lower plot shows using 93.5~dB.
The louder playback result\footnote{The LTI gain of Fig.~\ref{fig:smoothedOneTypResponse} and Fig.~\ref{fig:fostex10cm} are different because they were measured at a different time with different sensitivity setting.} shows that the signal-dependent distortion is significantly higher than the random response.
The signal-dependent random noise mentioned before contributes to this difference.


\section{Conclusion}
We proposed a simple method for measuring attributes of acoustic systems using arbitrary sounds by safeguarding them.
This method enables us to use music to measure room or concert acoustic conditions filled with the audience without annoying their musical experience.
This report is to demonstrate the feasibility of the proposed method.
This safeguarding makes even an arbitrary pure tone possible to measure the LTI response in the whole audible frequency range by adding a slight deterministic noise.

This method has a wide range of applications.
For example, for assessing the listening conditions of a classroom, we can use words and phrases.
We are planning theoretical and comprehensive investigations of this method and the introduction of frequency-dependent flooring.

\section*{Acknowledgement}
This work was supported by JSPS (Japan Society for the Promotion of Science) Grants-in-Aid for Scientific Research Grant Numbers JP21H00497 and JP20H00291.
The authors appreciate discussions and comments by Drs. Ken-Ichi Sakakibara, Tatsuya Kitamura, Mitsunori Mizumachi, and Hideki Banno.
The authors appreciate the loop music composition and performance very much by Kazuki Matsumoto.




\begin{thebibliography}{9} 
\bibitem{aoshima1981jasa} N. Aoshima, ``Computer-generated pulse signal applied for sound measurement,'' {\em JASA}, {\bf\sf69}(5), 1484--1488, (1980).
\bibitem{schroeder1970ieeeit} M. Schroeder, ``Synthesis of low-peak-factor signals and binary sequences with low autocorrelation,'' {\em IEEE Trans. Information Theory}, {\bf\sf16}(1), 85--89, (1970).
\bibitem{kawahara2020apsipa} H. Kawahara, K. -I. Sakakibara, M. Mizumachi, M. Morise and H. Banno, ``Simultaneous measurement of time-invariant linear and nonlinear, and random and extra responses using frequency domain variant of velvet noise,'' {\em Proc. APSIPA ASC 2020}, 174--183, (2020).
\bibitem{kawahara2021icassp} H. Kawahara and K. Yatabe, ``Cascaded All-Pass Filters with Randomized Center Frequencies and Phase Polarity for Acoustic and Speech Measurement and Data Augmentation,'' {\em Proc. ICASSP 2021}, 306--310, (2021).
\end{thebibliography}
\end{document}